\documentclass[pra,twocolumn,showpacs,floatfix]{revtex4}
\usepackage{graphicx}
\usepackage{color}
\usepackage{amsmath}
\begin{document}

\title{Rotational properties of superfluid Fermi-Bose mixtures in a tight toroidal trap}

\author{M. \"{O}gren$^{1,2}$ and G. M. Kavoulakis$^{2}$}

\affiliation{$^{1}$School of Science and Technology, \"{O}rebro University, 70182 \"{O}rebro, Sweden \\
$^2$Hellenic Mediterranean University, P.O. Box 1939, GR-71004, Heraklion, Greece}

\date{\today{}}

\begin{abstract}

We consider a mixture of a Bose-Einstein condensate, with a paired Fermi superfluid, confined in a 
ring potential. We start with the ground state of the two clouds, identifying the boundary between the 
regimes of their phase separation and phase coexistence. We then turn to the rotational response of the 
system. In the phase-separated regime, we have center of mass excitation. When the two species coexist, 
the spectrum has a rich structure, consisting of continuous and discontinuous phase transitions. 
Furthermore, for a reasonably large population imbalance it develops a clear quasi-periodic behaviour,
in addition to the one due to the periodic boundary conditions. It is then favourable for the one 
component to reside in a plane-wave state, with a homogeneous density distribution, and the problem 
resembles that of a single-component system.

\end{abstract}
 \maketitle

\section{Introduction}

One of the interesting achievements in the field of cold atomic gases is the realization of Fermi-Bose mixtures.
For example, in Ref.\,\cite{Schreck2001} the first observation of a mixture of a Bose-Einstein condensate in a 
Fermi sea was reported. An interesting possibility is that where both components are superfluids. Such experiments 
have been reported in a $^{6}$Li-$^{7}$Li mixture~\cite{Abeelen, Kempen, Ferrier, Delehaye}, in a $^{40}$K-$^{41}$K 
mixture with a tunable interaction between the two species~\cite{Wang, Falke, CWu}, in a mixture of $^{6}$Li and 
$^{133}$Cs with broad interspecies Feshbach resonances~\cite{Repp, Tung}, and in a two-component superfluid 
$^6$Li-$^{41}$K and $^6$Li-$^{174}$Yb~\cite{Yao,Roy}. 

In the problem of Fermi-Bose superfluid mixtures, various effects have been studied. These include Faraday 
waves~\cite{Abdullaev2013}, the existence of a super-counter-fluid phase~\cite{Kuklov}, the existence of 
dark-bright solitons~\cite{AdhikariPRA2007}, the multiple periodic domain formation~\cite{TylutkiNJP2016}, 
and collective oscillations ~\cite{Ferliano, Banerjee, Nascimbene, Wen, Wu, MitraJLTP2018, Abdullaev2018}. 
Finally, the ground state and the existence of vortices in a rotating quasi-two-dimensional Fermi-Bose mixtures 
has also been investigated in~\cite{WenPRA2014}.

Motivated by the studies mentioned above, in this article we study the ground state and the rotational properties 
of a mixture of a Bose superfluid, with a (paired) Fermi superfluid, at zero temperature. We consider the problem 
where the two components are confined in a ring potential, i.e., under the assumption of one-dimensional motion, 
and periodic boundary conditions. This situation is realized experimentally in a very tight toroidal potential, 
where the transverse degrees of freedom are frozen. We stress that this problem is not only interesting theoretically, 
but it is also experimentally relevant, following early experiment on rotating fermions in a harmonic 
trap \cite{Ketferm}. For example, numerous experiments have been performed on single-component rotating Bose-Einstein 
condensed atoms in annular/toroidal traps, see, e.g., Refs.\,\cite{ann1}. What is even more relevant is the experiment 
of Ref.\,\cite{ann2}, where a mixture of two distinguishable species of Bose-Einstein condensed atoms has been 
investigated.

One of the main results of the present study is the state of lowest energy for some fixed value of the angular 
momentum of this coupled system and the corresponding dispersion relation, which plays a central role in its 
rotational response. As shown below, this problem has a very rich and interesting structure. Given the large 
number of parameters, we have chosen to tune the relative strength of energy scales that are associated with 
the Bose-Bose coupling, the Bose-pair coupling, and the Fermi energy, which results from the fermionic origin 
of the pairs. 

An experimentally-relevant assumption which is done in the present study is that the kinetic energy associated 
with the motion of the atoms around the ring is much smaller than all these three energy scales. As a result, 
in the case of phase coexistence, it is energetically favourable for the system to reside in plane-wave states, 
with a homogeneous density distribution. This is simply due to the fact that, when the kinetic energy is 
negligible, the interaction energy is always minimized when the density of both components is homogeneous \cite{RoussouNJP2018}. Clearly, this simplifies the problem significantly, but on the same time it gives it 
an interesting structure, as the cloud undergoes discontinuous transitions as the angular momentum increases. 
In addition, the dispersion relation has a quasi-periodicity, which is set by the minority component. 

The system that we have considered is ideal for the study of one of the most fundamental problems 
in cold atomic systems, namely, superfluidity. One of the main messages of our study is the richness of the 
collection of phenomena which are associated with superfluidity due to the mixing of a Bose-Einstein 
condensate of bosonic atoms, with a paired superfluid Fermi system.

In what follows below, we start with our model in Sec.\,II. In Sec.\,III we derive the condition for phase 
coexistence and phase separation of the two superfluid components. We then turn to the rotational response 
of the system in Sec.\,IV. In this section we start with the phase-separated regime, and then turn to the 
case of phase coexistence. In Sec.\,V we investigate the more general problem where the boson mass is not 
equal to the mass of the pairs. Finally, we summarize our main results and present our conclusions in Sec.\,VI.

\section{The model} \label{sec_The_model}

The problem we have in mind is that of a fermionic and a bosonic superfluid, confined in a very tight 
toroidal trap. Following Ref.\,\cite{JMK}, since the transverse degrees of freedom are frozen, 
one may assume that the two order parameters of the bosonic atoms and of the fermionic pairs have a product 
form of a Gaussian profile in the transverse direction, times $\Psi_B(\theta)$ for the bosonic atoms and 
$\Psi_P(\theta)$ for the fermionic pairs, with both $\Psi_B$ and $\Psi_P$ depending on the azimuthal angle 
$\theta$, only. Then, integrating over the transverse direction, one ends up in the following energy functional,
\begin{eqnarray}
 H = - \frac {\hbar^2 N_P} {2 m_P R^2} \int \Psi_P^* \partial_{\theta \theta} \Psi_P \, d x
 - \frac {\hbar^2 N_B} {2 m_B R^2} \int \Psi_B^* \partial_{\theta \theta} \Psi_B \, d x
\nonumber \\ + \frac 1 3 G_P N_P^3 \int |\Psi_P|^6 \, d x
+ G_{FB} N_P N_B \int |\Psi_P|^2 |\Psi_B|^2 \, d x
\nonumber \\ + \frac 1 2 G_B N_B^2 \int |\Psi_B|^4 \, d x,
\nonumber \\
\label{Ham}
\end{eqnarray}
where $x = R \theta$, with $R$ being the radius of the ring. In the above equation we use the normalization 
$\int |\Psi_B|^2 d x = 1$ and $\int |\Psi_P|^2 dx = 1$. Here $N_B$ is the number of bosonic atoms, with a mass 
$m_B$. Assuming that we have $N_F$ number of fermionic atoms, divided equally into two spin states that pair up, 
we have $N_P = N_F/2$ pairs of fermions. Also, $m_P = 2 m_F$ is the mass of the fermion pairs, with $m_F$ being 
the mass of each fermionic atom. In the data presented below we first consider the case $m_B = m_P$, and then in 
Sec.\,V we examine the more general problem, $m_B \neq m_P$. 

Also, in Eq.\,(\ref{Ham}), $G_{B}$ and $G_{FB}$ are the intra- and inter- species one-dimensional interaction 
couplings~\cite{AdhikariPRA2007}, which are proportional to the scattering lengths $a_B$ and $a_{FB}$, respectively. 
Since in the quasi-one-dimensional scheme that we have adopted the transverse degrees of freedom of the order 
parameters are frozen, the form of the Hamiltonian of Eq.\,(\ref{Ham}) is essentially the same as the one of 
the purely one-dimensional problem. The last term in $H$, which is proportional to $G_B$, is the usual quartic 
term, which corresponds to the boson-boson interaction. This is the leading-order term of the more general expression 
derived by Lieb and Liniger \cite{LL}, for weak boson-boson coupling. The term which is proportional to $G_{FB}$, is 
similar to the last one and it describes the interaction between the bosons and the pairs, with the only difference 
being that it is proportional to the product of the densities of the bosons and of the pairs. Finally, the term which 
is proportional to $G_P$ comes from the Fermi energy of the fermionic atoms which constitute the pairs. This is also 
the leading-order term of a more general (Gaudin-Yang) Hamiltonian \cite{GY}, valid both in the BCS and in the
molecular limits.

The standard formula that connects $G_B$ with the scattering length $a_B$ is $G_{B}=2\hbar \omega_{\perp,B} 
a_{B}$. Here, $\omega_{\perp,B}$ is the frequency of the trap in the transverse direction. The formula for 
$G_{FB}$ is more complicated, since it depends on the atomic masses and, potentially, on the two different 
trap frequencies. Here we choose $G_{FB}$ -- compared with $G_B$ -- in such a way that we explore the 
physically-interesting regimes. Finally, $G_P = 4 \kappa \hbar^2 \pi^2/m_F$, with the dimensionless parameter 
$\kappa = 1/4$ in the BCS, weakly-attractive coupling limit, and $\kappa = 1/16$ in the molecular unitarity 
limit~\cite{AdhikariPRA2007}, i.e., close to a Feshbach resonance associated with the fermion-fermion 
interaction. In all the results which follow below $\kappa$ is set equal to $1/4$, however 
they are relatively insensitive to this value, in the sense that the results are not affected, at least 
qualitatively. Clearly $G_B$ and $G_{FB}$ have the same units, i.e., energy times length, while $G_P$ has 
units of energy times length squared.

In order for the Hamiltonian of Eq.\,(\ref{Ham}) to be valid, the bosonic atoms have to be in 
the mean-field regime, which means that their density per unit length $n_B^0 = N_B/(2 \pi R)$ must be 
larger than $\approx a_B/a_{\perp,B}^2$, where $a_{\perp,B}$ is the oscillator length that corresponds to 
$\omega_{\perp,B}$. Furthermore, the assumption of quasi-one-dimensional motion requires that $n_B^0 a_B 
\ll 1$. Therefore, $n_B^0$ has to be in the range $a_B/a_{\perp,B}^2 \ll n_B^0 \ll 1/a_B$. For the 
fermionic component, the third term in Eq.\,(\ref{Ham}) is valid in both limits of weak $(\kappa = 1/4)$ 
and strong $(\kappa = 1/16)$ attraction.

From Eq.\,(\ref{Ham}) it follows that in the rotating frame with some angular frequency $\Omega$ \cite{LLb}, 
$\Psi_P(\theta)$ and $\Psi_B(\theta)$ satisfy the two coupled equations \cite{AdhikariPRA2007, WenPRA2014}, 
\begin{widetext}
\begin{eqnarray}
 \left[ -\frac{\hbar^2}{2 m_B R^2}  \partial_{\theta \theta} 
 + i \Omega \hbar \partial_{\theta} +  G_B N_B |\Psi_B|^2 +  G_{FB} N_P |\Psi_P|^2  \right] \Psi_B  &=& \mu_B \Psi_{B} \; , 
 \nonumber \\
 \left[ -\frac{\hbar^2}{2 m_P R^2} \partial_{\theta \theta} + i \Omega \hbar \partial_{\theta} 
 + G_P N_P^2 |\Psi_P|^4 +  G_{FB} N_B |\Psi_B|^2 \right] \Psi_P  &=& \mu_P \Psi_{P} .  
 \label{sys1}
\end{eqnarray}
\end{widetext}
Alternatively, we can view Eqs.~(\ref{sys1}) as the Euler-Lagrange equations for the minimization of the total 
energy, under the following three constraints: a constant total angular momentum $-i \hbar R \int [N_B \Psi_B^* 
\partial_{\theta} \Psi_B + N_P \Psi_P^* \partial_{\theta} \Psi_P] d \theta = L_B \hbar + L_P \hbar = L \hbar$, 
and a fixed number of $N_B$ bosonic atoms and $N_P$ fermionic pairs, with the three corresponding Lagrange 
multipliers being $\Omega$, $\mu_B$ and $\mu_P$. In the results that follow below we have thus fixed $L$
and we have evaluated the state of lowest energy, treating $\Omega$ as a Lagrange multiplier. 

We stress that in a harmonic trap, in two, or three spatial dimensions there is the following possibility: 
When rotated, a paired Fermi system, which is in the BCS regime (only) may form a shell of unpaired atoms, 
which undergo solid-body rotation, along with a core of non-rotating paired atoms, which are located near 
the center. As argued in Ref.\,\cite{str}, in this case breaking of the pairs is energetically inexpensive, 
since the density is low, while the cloud gains energy due to the centrifugal energy of the normal cloud, 
undergoing solid-body rotation. In the present problem such a ``decoupling" between the pair and the 
unpaired parts of the cloud is not possible, since the two parts would have to move together. Therefore, 
we do not expect such an effect to be present here (which would anyway be relevant for $\kappa \approx 1/4$, 
only).  

Finally, we should mention that in our model we have excluded the possibility of Efimov states \cite{ES}. 
Whether these play any role in our problem would be the subject of a separate publication. Such a study 
should also investigate the effect of losses.

\section{Boundary between phase separation and phase coexistence of the two superfluid components} 
\label{Sec_Mixing}

Depending on the value of the parameters, there are two phases. In the one, the two components have 
an inhomogeneous density distribution and prefer to reside in different parts of the torus/ring, i.e., 
we have phase separation. In the other, the two components have a homogeneous density distribution, 
and thus we have phase coexistence. We derive the condition for energetic stability of the homogeneous 
phase in Appendix~\ref{app_sec_for_mixing_condition} and the condition for its dynamic stability from 
the Bogoliubov spectrum, in Appendix B. 

The condition for energetic stability of the phase where the two components are distributed homogeneously 
and coexist is
\begin{widetext}
\begin{eqnarray}
\left( \frac{\hbar^2}{2 m_B R^2} + 2 G_B n_B^0 \right) 
\left( \frac{\hbar^2}{2 m_P R^2} + 4 G_P (n_P^0)^2  \right) > 4 G_{FB}^2 n_B^0 n_P^0,
\label{CriticalMixingDemixing}
\end{eqnarray}
\end{widetext}
where $n_B^0 = N_B/(2 \pi R)$ and $n_P^0 = N_P/(2 \pi R)$. Let us denote as $\phi_k = e^{i k \theta}/\sqrt{{2\pi R}}$, 
with $k$ being an integer, the well-known eigensolutions of the (single-particle) kinetic-energy operator $- \hbar^2 
\partial_{\theta \theta}/(2 m_P R^2) = - \hbar^2 \partial_{\theta \theta}/(2 m_B R^2)$, that corresponds to the kinetic 
energy of the particles, under periodic boundary conditions, with an energy $\hbar^2 k^2/(2 m_P R^2) = \hbar^2 k^2/(2 m_B 
R^2)$ and angular momentum $k \hbar$. It is clear that if Eq.\,(\ref{CriticalMixingDemixing}) is satisfied for the 
non-rotating state ($L=0$), where $\Psi_B = \phi_0$ and $\Psi_P = \phi_0$, it will also be satisfied for any plane-wave 
state with nonzero angular momentum $(k_B,k_P)$, where 
\begin{equation}
(k_B,k_P) \equiv \left( \Psi_B = \phi_{k_B} , \ \Psi_P = \phi_{k_P} \right).  
\label{eq_PW}
\end{equation}
We stress that this analysis is local and not global. In other words, the condition of Eq.\,(\ref{CriticalMixingDemixing}) 
does not necessarily imply that any state $(k_B,k_P)$ is the absolute minimum of the energy for the given angular momentum 
per particle $\ell = L/N = x_B k_B + x_P k_P$, where $N = N_B + N_P$. Still, it becomes global when the nonlinear terms 
in the Hamiltonian are sufficiently large \cite{Zar}, or, equivalently, when the kinetic energy is sufficiently small 
(as in the present problem).

In addition to the energetic stability examined above there is also the dynamic stability of the homogeneous solution.
In Appendix B we derive the Bogoliubov spectrum, i.e., the excitation energy $\hbar \omega(k)$, as function of $k$, 
where $k \hbar$ is the angular momentum of the system, with $k$ being an integer. This is given by the smaller value 
of $\omega^2$ which solves the following equation,
\begin{widetext}
\begin{eqnarray}
  \left( \frac {\hbar^2 k^4} {2 m_B R^2} + 2 G_B n_B^0 k^2 - 2 m_B \omega^2 R^2 \right) 
  \left( \frac {\hbar^2 k^4} {2 m_P R^2} + 4 G_P (n_P^0)^2 k^2 - 2 m_P \omega^2 R^2 \right) =
 4 G_{FB}^2 n_B^0 n_P^0 k^4.
 \label{ds}
\end{eqnarray}
\end{widetext}
The condition for dynamic stability which results from Eq.\,(\ref{ds}) with $k=1$, i.e., for real $\omega$, coincides 
with that of energetic stability, Eq.\,(\ref{CriticalMixingDemixing}). In addition, it follows from Eq.\,(\ref{ds}) 
that in the case of a purely bosonic component
\begin{eqnarray}
  m_B c_B^2 = \frac {\hbar^2} {4 m_B R^2} + G_B n_B^0,
\end{eqnarray}
where $c_B$ is the bosonic speed of sound. Similarly, in a purely fermionic superfluid,
\begin{eqnarray}
  m_P c_P^2 = \frac {\hbar^2} {4 m_P R^2} + 2 G_P (n_P^0)^2,
\end{eqnarray}
where $c_P$ is the speed of sound of the fermionic pairs. 

\section{Rotational response of the two superfluids} \label{Sec_NumYrastStates}

Let us now turn to the rotational response of the system, under some fixed angular momentum, that we examine below. 
This problem depends on whether the two components are separated, or they coexist. For this reason, we examine each 
case separately, below. 

Before we proceed, it is instructive to identify the three energy scales $E_B$, $E_{FB}$, and $E_F$, which appear 
in the Hamiltonian of Eq.\,(\ref{Ham}), defined as
\begin{eqnarray}
\frac {E_B} N = 
 \frac 1 2 x_B^2 G_B n_0 , \,\, 
\frac {E_{FB}} N = 
 x_B x_P G_{FB} n_0, 
 \label{ensc0}
\end{eqnarray}
and
\begin{eqnarray}
\frac {E_F} N = \frac 1 3 G_P x_P^3 n_0^2, 
\label{ensc} 
\end{eqnarray}
where $n_0 = N/(2 \pi R)$, $x_P = N_P/N$, and $x_B = N_B/N$. Denoting as $K$ the kinetic energy per particle, 
$K = \hbar^2/(2 m_B R^2) = \hbar^2/(2 m_P R^2)$, we introduce the useful dimensionless quantities $\epsilon_B = 
{E_B}/(N K)$, $\epsilon_{FB} = {E_{FB}}/(N K)$, and $\epsilon_F = {E_F}/(N K)$. In what follows below we consider 
values of $\epsilon_B$, $\epsilon_{FB}$, and $\epsilon_F$ which are much larger than unity, as is also the case 
experimentally. 

The terms $E_B$ and $E_{FB}$ are the familiar ones, met also in the case of boson-boson mixtures. On the other 
hand, $E_F$ comes from the Fermi pressure of the ``underlying" fermionic origin of the pairs, and it acts as an 
effective repulsive potential, which, however, does not scale with the density in the usual, quadratic, way, that 
we are familiar with from the case of contact interactions. In this case of the pairs, the corresponding energy 
per unit length increases with the third power of the pair density, i.e., there is a stronger dependence of this
term on the density. This is one of the major differences between the problem of Fermi-Bose, and Bose-Bose mixtures. 
Finally, we remark that while the contact potential corresponds to two-body collisions, $E_F$ resembles a term that 
corresponds to three-body collisions. 

Before we proceed, we stress that Bloch's theorem \cite{FB}, which refers to a single component system, is also 
valid in our two-component system, at least under certain conditions, which are examined in Sec.\,V \cite{prl, 
Anoshkin}. The easiest case is that of equal masses, $m_B = m_P$, where the energy is a periodic function, on 
top of a parabola, i.e.,
\begin{eqnarray}
  \frac E {N K} = \ell^2 + \frac {e(\ell)} K.
  \label{bl}
\end{eqnarray}
Here $e(\ell)$ is a periodic function, with a period equal to unity, $e(\ell + 1) = e(\ell)$. In what follows 
below we measure the zero of the energy with respect to the ground-state energy of the nonrotating system, and
therefore $e(\ell = 0) = 0$ in what follows below. 

\subsection{Regime of phase separation}

\begin{figure}
\includegraphics[scale=0.6]{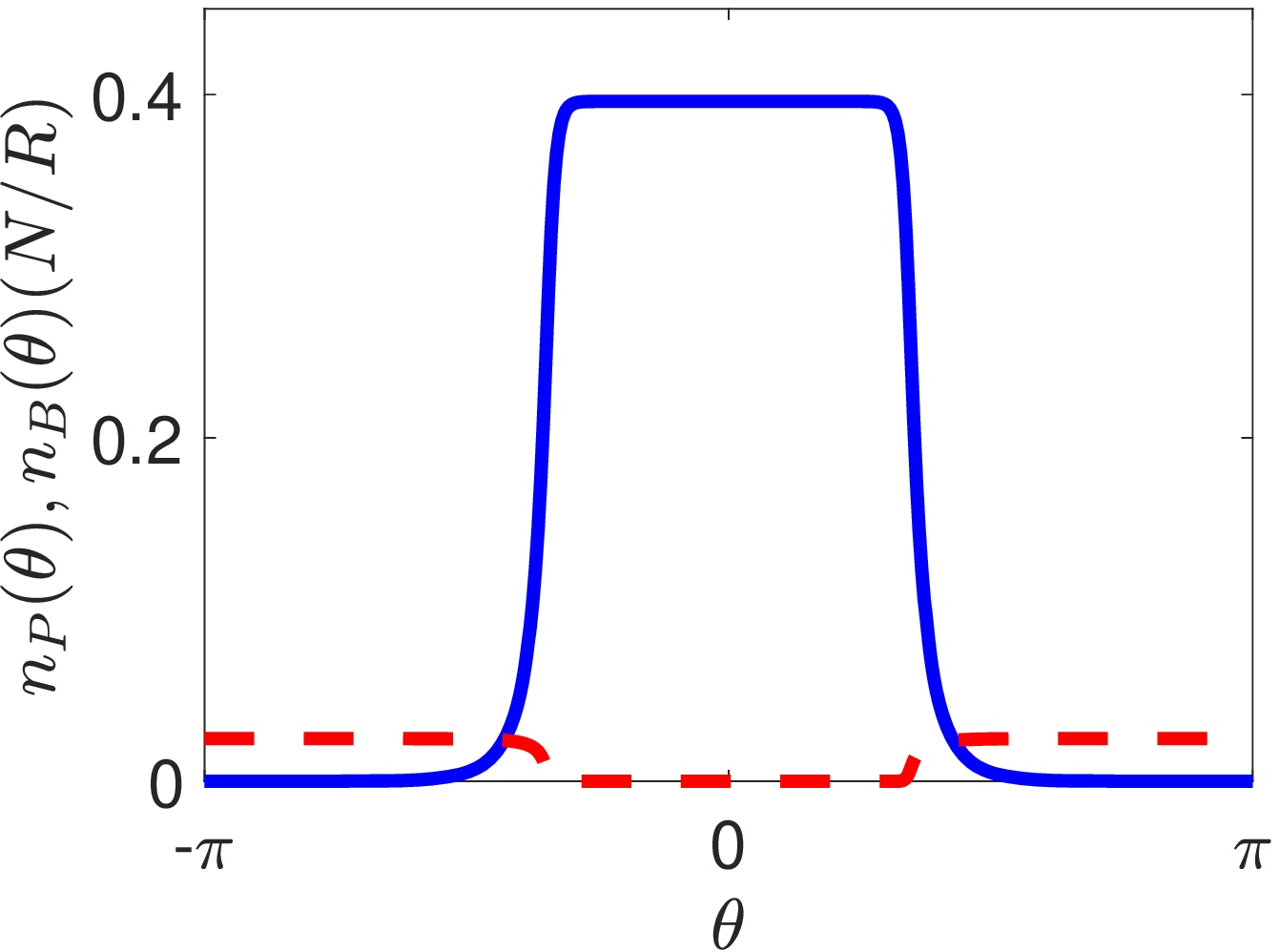}
\includegraphics[scale=0.6]{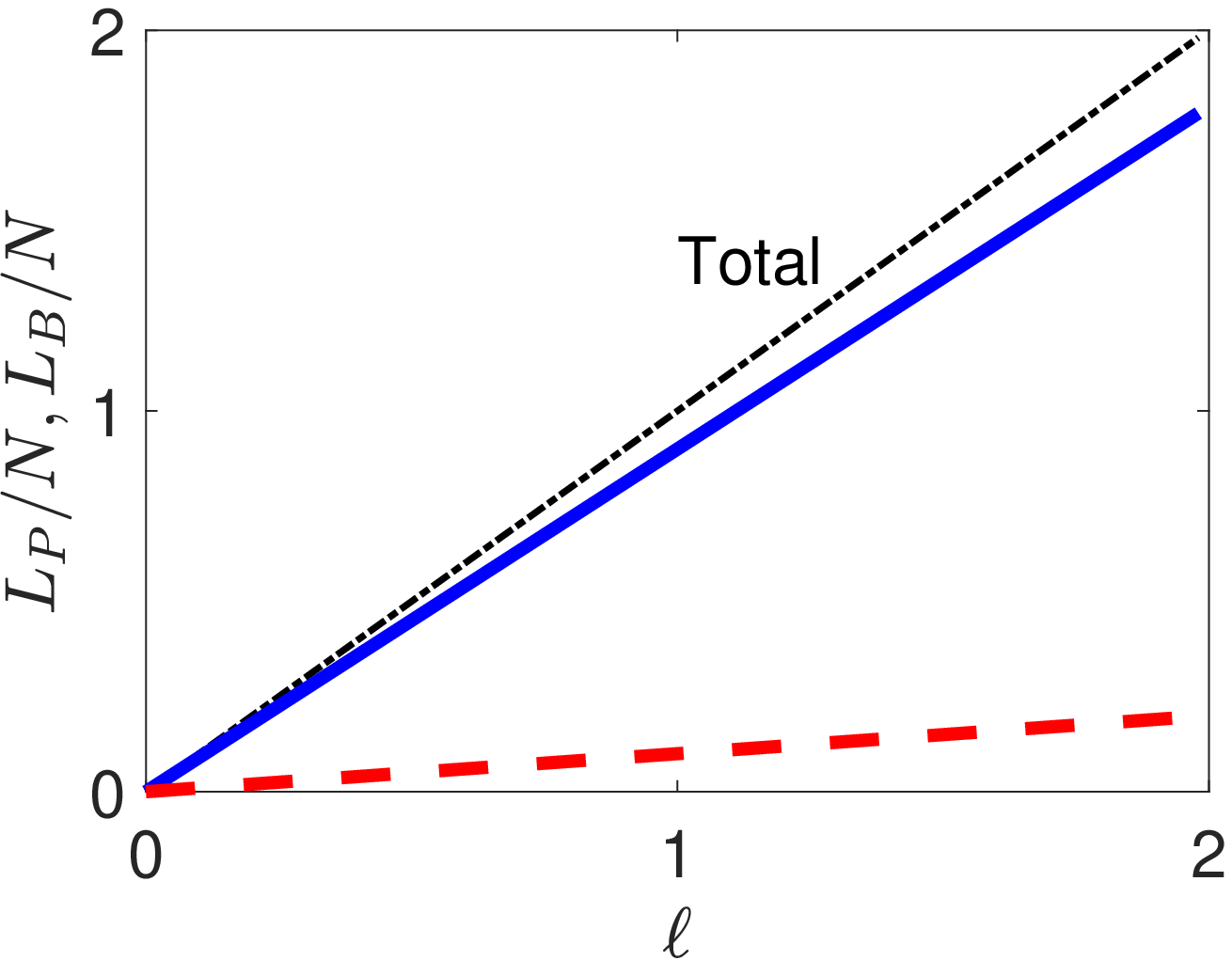}
\caption{(Color online) The density distribution $n_P(\theta) = N_P |\Psi_P(\theta)|^2$ and $n_B(\theta) 
= N_B |\Psi_B(\theta)|^2$ of the two components (upper plot), and the distribution of the angular momentum 
between the two components (lower plot), in the regime of phase separation. Here the blue color (solid line) 
corresponds to the bosons, and the red color (dashed line) corresponds to the pairs. Also, $x_P=0.1$ and 
$x_B=0.9$, in the BCS regime ($\kappa = 1/4$), $\epsilon_B \approx 83.33$, $\epsilon_{FB} \approx 300.0$, 
and $\epsilon_F \approx 83.33$. The density of the two components is independent of the angular momentum. 
On the lower figure, $\ell_{B,P}/\ell = m_{B,P} N_{B,P}/(m_B N_B + m_P N_P)$, as explained in the text.}    
\label{fig:Demixing}
\end{figure}

In the regime of phase separation, already for zero angular momentum the density of the two components is 
inhomogeneous. In Fig.\,\ref{fig:Demixing} we show numerical solutions of Eqs.\,(\ref{sys1}), i.e., the 
results we have derived minimizing the energy of the system, fixing the angular momentum. More specifically, 
we show the density distribution of the two components, and the angular momentum carried by each component 
separately. In these results we choose a large enough value of $G_{FB}$, see Eq.\,(\ref{CriticalMixingDemixing}), 
so that the demixing is complete, i.e., the two densities have almost zero overlap, see the upper plot of 
Fig.~\ref{fig:Demixing}. In this case it is energetically favourable for the system to carry the angular 
momentum via center of mass excitation. 

In Fig.\,\ref{fig:Demixing} we also consider a population imbalance $x_P = N_P/N = 0.1$ and $x_B = N_B/N = 0.9$, 
while $\epsilon_B \approx 83.33$, $\epsilon_{FB} \approx 300.0$, and $\epsilon_F \approx 83.33$. From the lower 
plot of Fig.\,1 we see that the angular momentum is shared by the two components in a trivial way. More specifically, 
the total angular momentum $L \hbar = L_P \hbar + L_B \hbar$ is divided into the two components proportionately to 
the mass and the particle number, that is
\begin{equation}
\frac {L_P} L = \frac{m_P N_P}{m_B N_B + m_P N_P}, \, \frac {L_B} L = \frac{m_B N_B} {m_B N_B + m_P N_P}.
\label{share}
\end{equation}
Since the rotational kinetic energy $K_r$ of the two components is
\begin{equation}
K_r = K_{r,P} + K_{r,B} = \frac {\hbar^2 L_P^2} {2 N_P m_P R^2} + \frac {\hbar^2 L_B^2} {2 N_B m_B R^2},
\end{equation}
if follows trivially from Eq.\,(\ref{share}) that
\begin{eqnarray}
K_r = \frac {\hbar^2 L^2} {2 (m_B N_B + m_P N_P) R^2}.
\label{kr}
\end{eqnarray}
Therefore, the dispersion relation is exactly parabolic, in agreement with our numerical results (this is
also consistent with Bloch's theorem). Finally, we stress that the density distribution of the two components 
is unaffected by the rotation, since the system is excited via center of mass excitation. As a result, the 
density distribution shown in the upper plot of Fig.\,1 is independent of $\ell = L/N$. 

\subsection{Regime of phase coexistence}

We now turn to the problem where the two components have a homogeneous density distribution and thus they coexist. 
In this case the solution of Eqs.~(\ref{sys1}) for the (non-rotating) ground state is the trivial one, i.e., the 
plane-wave state $\Psi_B = \phi_0$, and $\Psi_P = \phi_0$.

\begin{figure}
\includegraphics[scale=0.6]{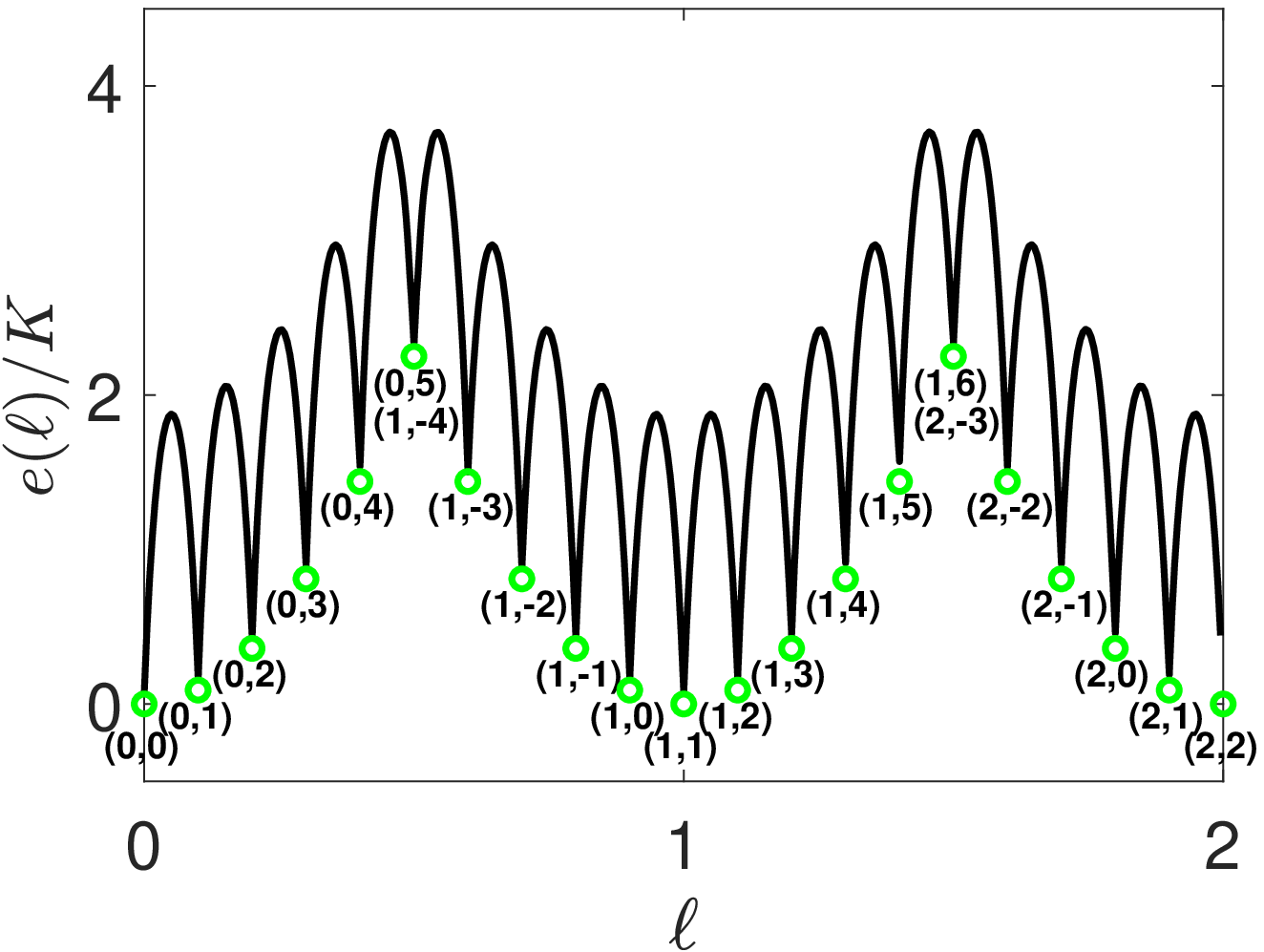}
\includegraphics[scale=0.6]{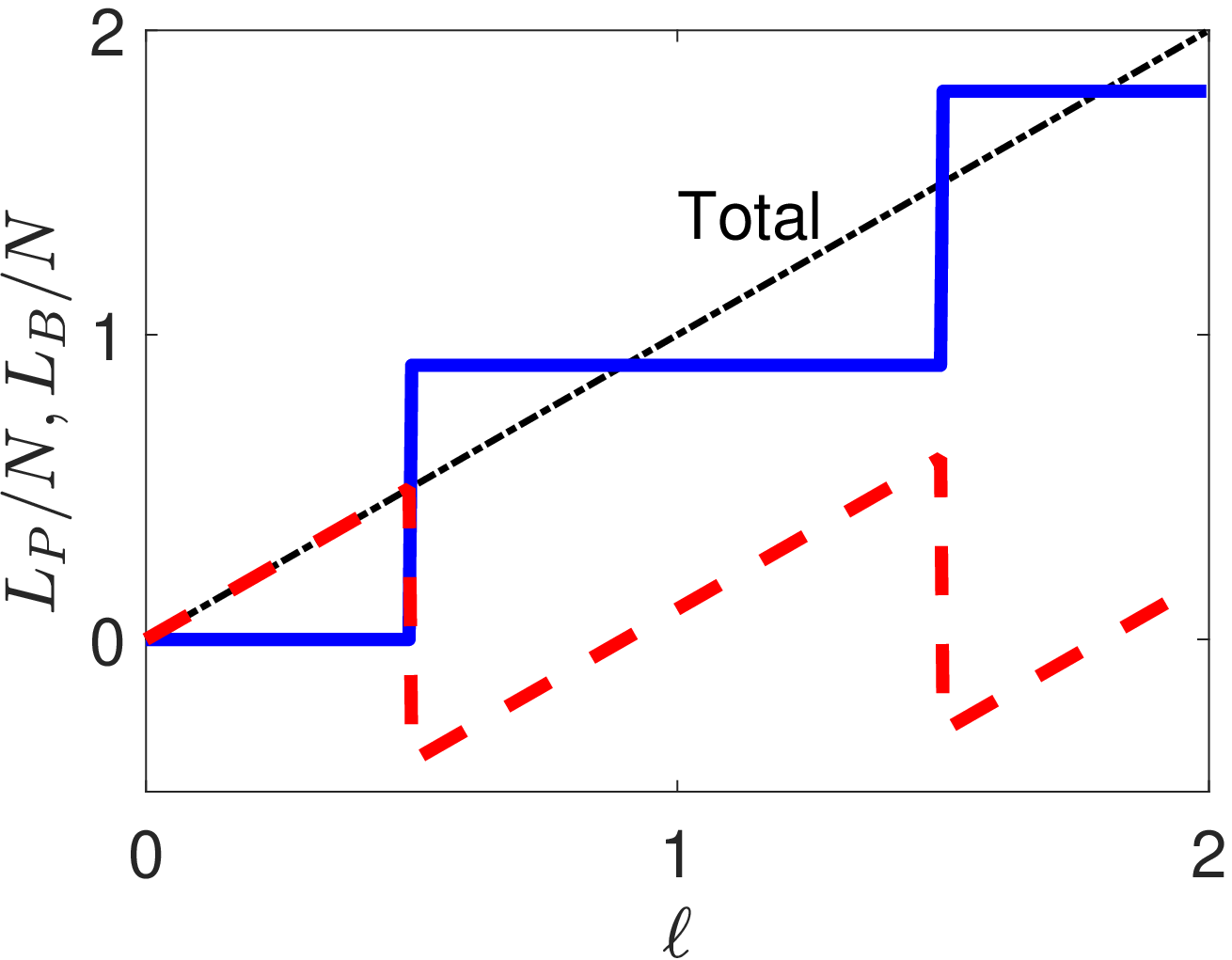}
\caption{(Color online) The periodic function $e(\ell)$ (upper plot), and the distribution of the angular 
momentum between the two components (lower plot), in the regime of phase coexistence. Here the blue color 
(solid line) corresponds to the bosons, and the red color (dashed line) corresponds to the pairs. Also, 
$x_P=0.1$ and $x_B = 0.9$, in the BCS regime ($\kappa=1/4$), $\epsilon_B \approx 3223$, $\epsilon_{FB} 
\approx 143.2$, and $\epsilon_F \approx 83.33$. The quasi-periodic behaviour in $L/N$ is set by $N_P/N 
= 0.1$. On the lower figure we see that the fermionic pairs carry (almost) all the angular momentum up 
to $\ell = 1/2$, since $\Psi_B \simeq \phi_0$ at this interval. For $1/2 < \ell < 3/2 $ the bosonic 
order parameter $\Psi_B$ is $\simeq \phi_1$. Finally, for $3/2 < \ell < 5/2$, $\Psi_B \simeq \phi_2$, 
etc. The indices, e.g., $(0,0)$ denote the order parameter of the two superfluids at each local minimum, 
in the notation of Eq.\,(\ref{eq_PW}).} 
\label{fig:3}
\end{figure}

\begin{figure}
\includegraphics[scale=0.6]{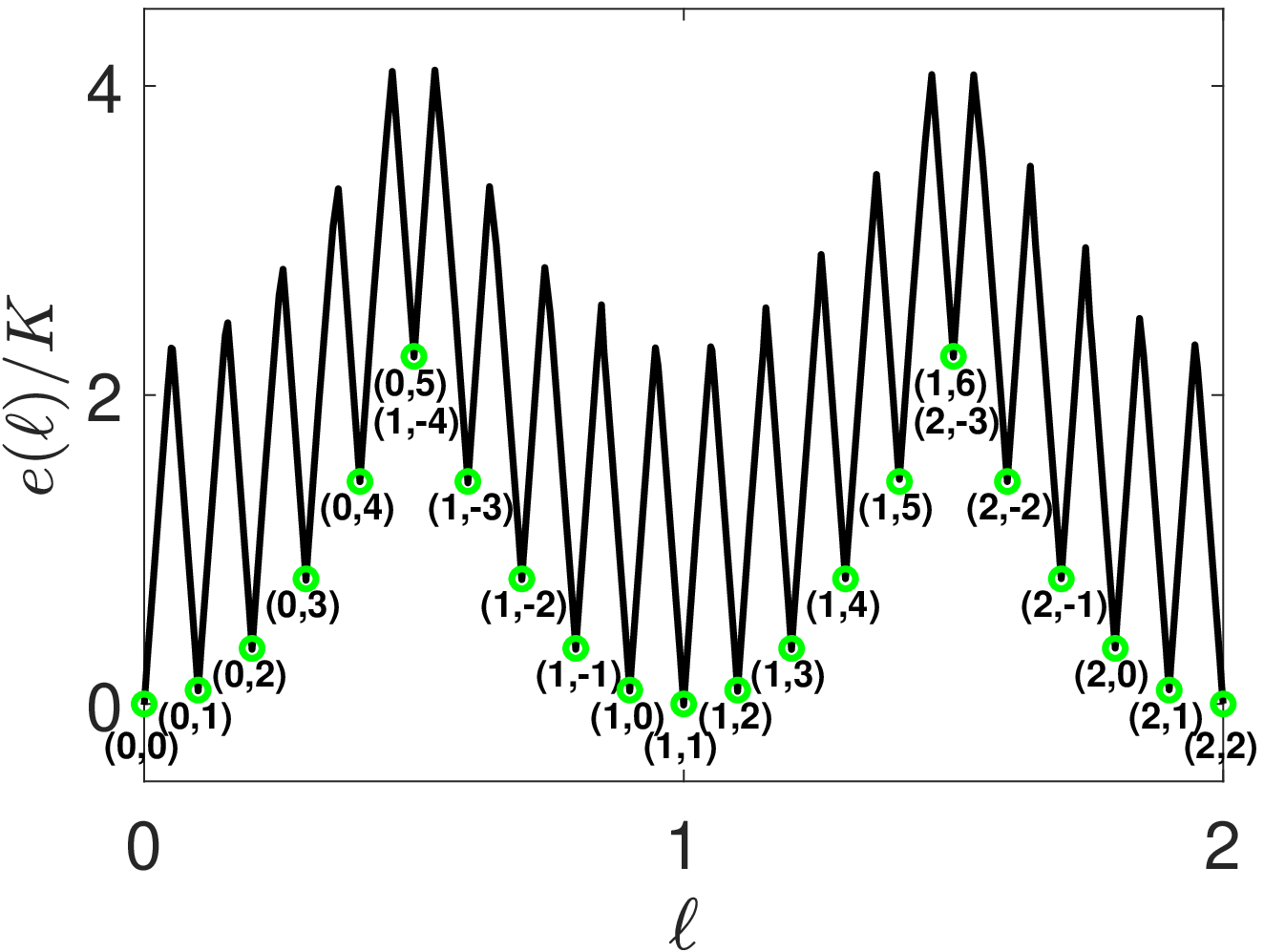}
\includegraphics[scale=0.6]{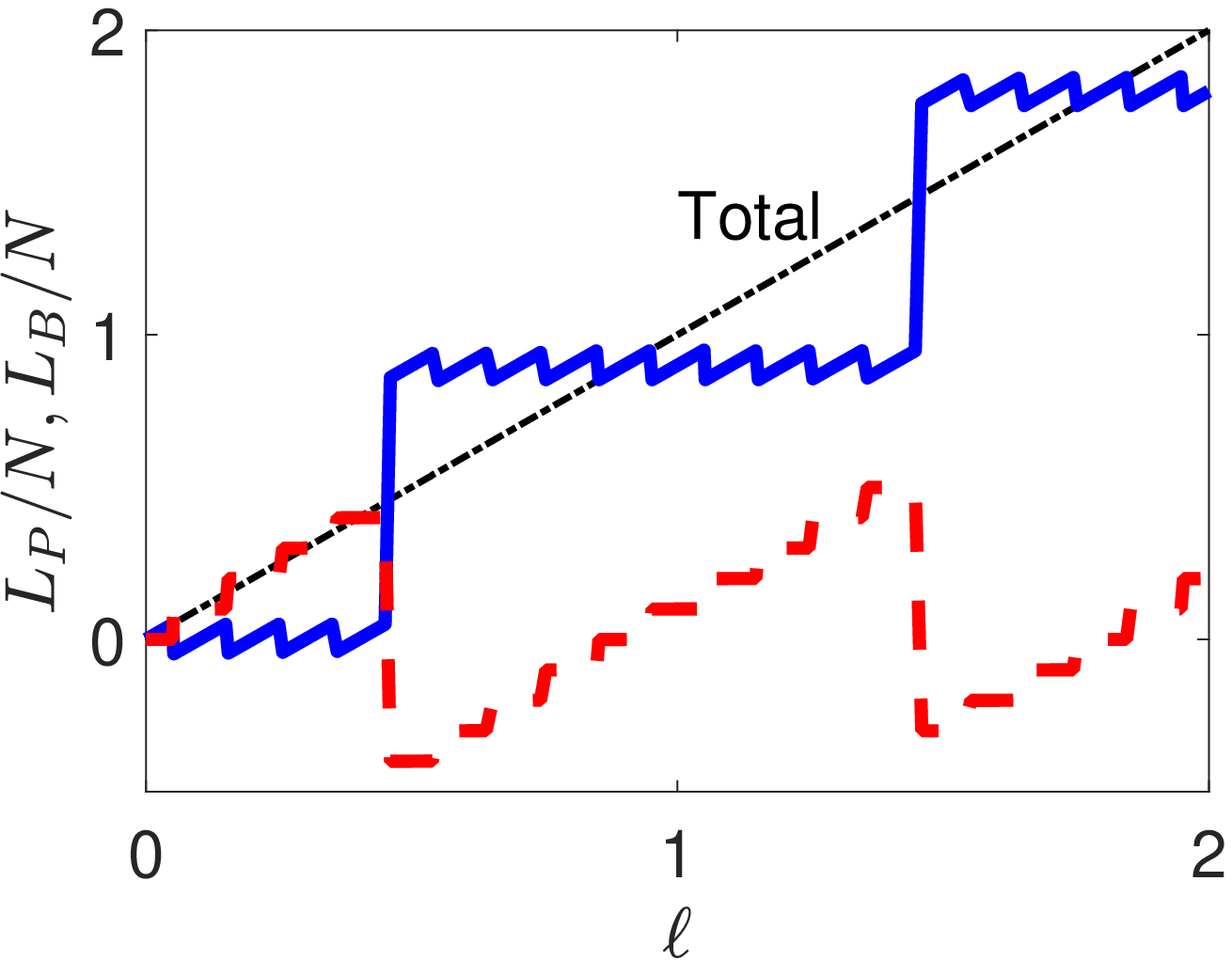} 
\caption{(Color online) The periodic function $e(\ell)$ (upper plot), and the distribution of the angular 
momentum between the two components (lower plot), in the regime of phase coexistence. Here the blue color 
(solid line) corresponds to the bosons, and the red color (dashed line) corresponds to the pairs. Also, 
$x_P=0.1$ and $x_B = 0.9$, in the BCS regime ($\kappa=1/4$), $\epsilon_B \approx 644.6$, $\epsilon_{FB} 
\approx 143.2$, and $\epsilon_F \approx 83.33$. The pairs are now in plane-wave states (as opposed to 
Fig.~\ref{fig:3}). The indices, e.g., $(0,0)$ denote the order parameter of the two superfluids at each 
local minimum, in the notation of Eq.\,(\ref{eq_PW}).}  
\label{fig:5} 
\end{figure}

In Figs.\,\ref{fig:3} and \ref{fig:5} we show again numerical solutions of Eqs.\,(\ref{sys1}), i.e., the results 
we have derived minimizing the energy of the system, fixing the angular momentum to some values. In these data we 
consider a population imbalance of $x_P = N_P/N = 0.1$ and $x_B = N_B/N = 0.9$, in the BCS regime ($\kappa=1/4$). 
Finally, we choose $\epsilon_{FB} \approx 143.2$ for the Fermi-Bose coupling, $\epsilon_F \approx 83.33$ and two 
different values for the Bose-Bose coupling, $\epsilon_B \approx 3223$ in Fig.\,\ref{fig:3}, and $\epsilon_B \approx 
644.6$ in Fig.\,\ref{fig:5}. 

Before we discuss each graph separately, let us start with some general remarks. From Eq.\,(\ref{bl}), subtracting 
from the energy of the system the energy due to the center of mass excitation, $E/(NK) - \ell^2$, we get the function 
$e(\ell)/K$, which is shown in the upper plot of Fig.\,2 and Fig.\,3. This has an exact periodicity of $\ell = 1$, 
due to Bloch's theorem. On top of this, it also has a quasi-periodicity, $\ell = 0.1$. As we show below, this is set 
by the minority component, and for the parameters we have chosen this is the paired fermionic superfluid, with
$x_P = N_P/N = 0.1$. 

Furthermore, when $\ell = L/N$ is an integer multiple of $x_P = N_P/N$, in their lowest energy, the two components 
are always in plane-wave states $(k_B, k_P)$, due to the large value of $\epsilon_B, \epsilon_{FB}$, and $\epsilon_F$ 
that we have considered \cite{RoussouNJP2018, Zar}. The actual value of $k_B$ and $k_P$ is determined by the 
minimization of the corresponding kinetic energy, $x_B k_B^2 + x_P k_P^2$, under the constraint of the fixed 
angular momentum, $\ell = x_B k_B + x_P k_P$, that we want the system to have. Clearly this does not depend on 
the value of the rest of the parameters, and for this reason the value of $k_B$ and $k_P$ is the same (for some 
given value of $\ell$) in both figures.

When $\epsilon_B$ and $\epsilon_F$ are much larger than unity, and we have phase coexistence, it is energetically 
favourable for the system to have a homogeneous density distribution in both components, for any value of $\ell$
\cite{RoussouNJP2018}. With the constraint of angular momentum this is not always possible, though. For small
values of the angular momentum, where we have sound waves (and the dispersion relation is linear in $\ell$), there 
is predominantly excitation of only one of the components. This may be seen from the (smaller) solution for $\omega R$ 
of Eq.\,(\ref{ds}), which is the speed of sound, and the corresponding eigenvector of the matrix of 
Eq.\,(\ref{matrix}). 

\begin{figure}
\includegraphics[scale=0.6]{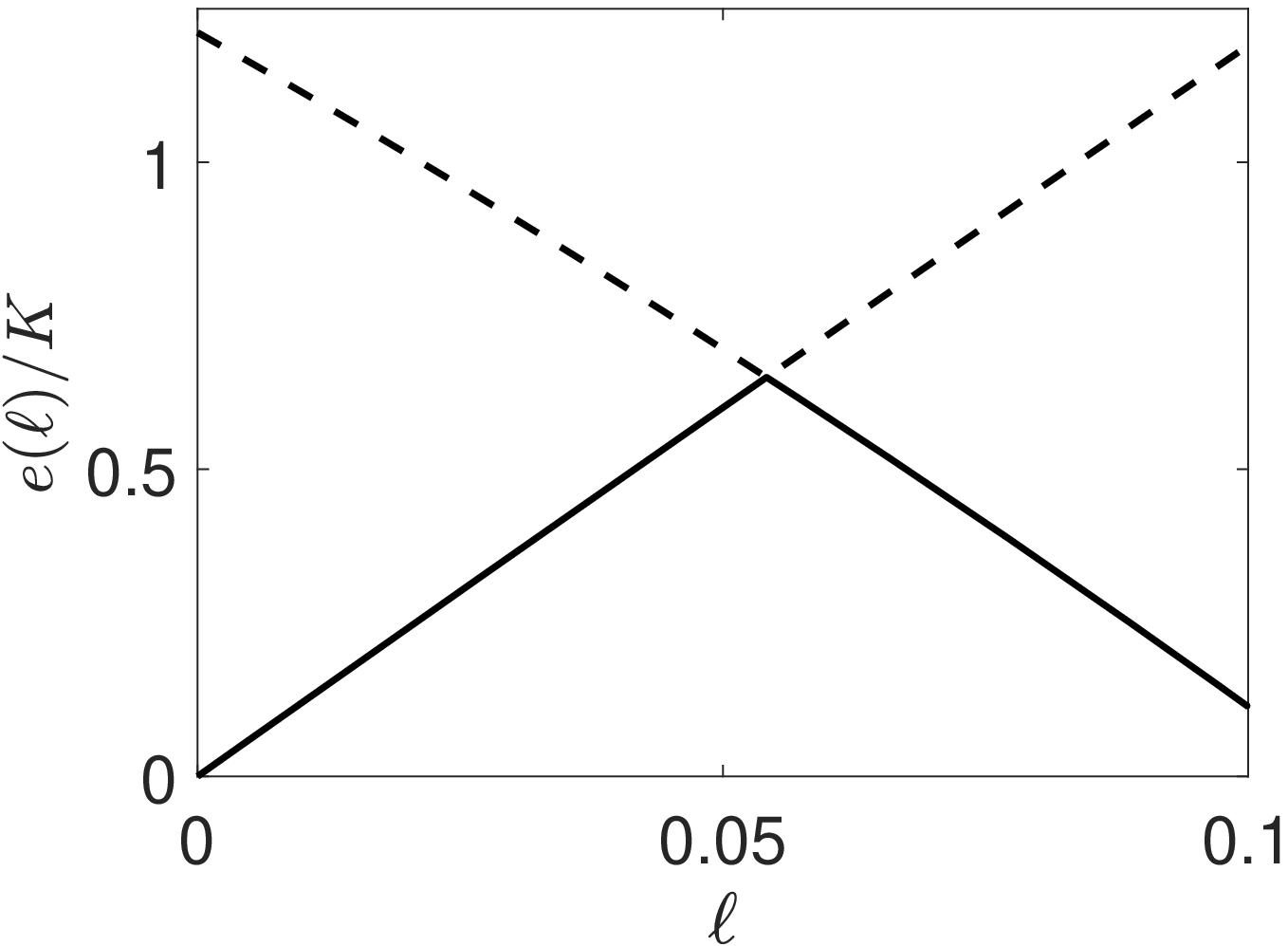}
\vspace{1cm}
\caption{The periodic function $e(\ell)$, derived from the two parabolas of Eqs.\,(\ref{E1}) and (\ref{E2}), 
for $\epsilon_B = 5$. Here $x_P = 0.1$ and $x_B = 0.9$. There is a level crossing at $\ell_0 \approx x_P/2 = 
0.05$. The system follows the lower, solid, curves, undergoing a discontinuous phase at the crossing point, 
i.e., at $\ell = \ell_0$.
\label{fig:cross}}
\end{figure}

For the chosen parameters, we evaluate from Eq.\,(\ref{ds}) the speed of sound, i.e., the slope of the dispersion
relation to be $\approx c_P \approx 50.0 \, \hbar/(m_P R)$ in Fig.\,\ref{fig:3}, in agreement with our numerical data. 
In this case we have (predominantly) excitation of the fermion pairs. In Fig.\,\ref{fig:5} the speed of sound is 
$\approx c_B \approx 26.17 \, \hbar/(m_B R)$ -- again in agreement with our numerical results -- we have (predominantly) 
excitation of the bosonic component. These results already help us explain Figs.\,\ref{fig:3} and \ref{fig:5}, at 
least for sufficiently small values of $\ell$.

Let us examine each plot separately, now, starting with Fig.\,\ref{fig:3}. In this case it is energetically 
favourable for the bosonic component to reside always in plane-wave states and thus retain its homogeneous 
density distribution, for all values of the angular momentum. In that sense, the ``interesting" component is 
the fermionic in this case. As a result, the density of the pairs changes as the angular momentum is varied, 
having the usual density distribution of solitary-wave excitation under periodic boundary conditions \cite{solf}, 
while the bosonic density is (to a rather good precision) constant. 

For $0 \le \ell < 1/2$, $\Psi_B \simeq \phi_0$ and the fermionic pairs carry (almost) all the angular momentum (as 
seen in the lower plot of Fig.\,2). Within this interval, we observe the subintervals with a width of $\Delta \ell 
= N_P/N = x_P = 0.1$ (as seen in the upper plot). For $0 < \ell < x_P$ the system shows all the characteristics of 
an ``ordinary" solitary wave (in the pairs), with a dispersion relation that has a negative curvature. This is 
due to the effective repulsive potential, discussed above (see Eq.\,(\ref{ensc}) and the discussion that follows 
below it). 

In order to get some insight, it is instructive to write down the trial (two-state) order parameter for the pairs 
in the limit of ``weak" interactions, i.e., when the three energy scales of Eqs.\,(\ref{ensc0}) and (\ref{ensc}) 
are much smaller than the kinetic energy $K$. Then, for $0 < \ell < x_P$,
\begin{eqnarray}
  \Psi_B = \phi_0, \,\, \Psi_P = \sqrt{1 - \frac {\ell} {x_P}} \phi_0 + \sqrt{\frac {\ell} {x_P}} \phi_1.
\end{eqnarray} 

At the second subinterval, $x_P < \ell < 2 x_P$, $\Psi_P$ changes trivially. More specifically, we have center 
of mass excitation, and thus $\Psi_P$ is simply multiplied by the phase $e^{i \theta}$. We stress that this 
does not affect the interaction energy, altering only the kinetic energy (and, obviously, the angular momentum).
Therefore, 
\begin{eqnarray}
  \Psi_B = \phi_0, \,\, \Psi_P = \sqrt{1 - \frac {\tilde{\ell}} {x_P}} \phi_1 + \sqrt{\frac {\tilde{\ell}} {x_P}} \phi_2,
\end{eqnarray}
where $0 < \tilde{\ell} < x_P$. This continues all the way, up to the interval $4 x_P < \ell < 5 x_P = 1/2$. 

For $1/2 < \ell < 6 x_P$, instead of the pair order parameter $\Psi_P$ to be multiplied by the phase $e^{5 i \theta}$, 
it is more favourable for the bosonic order parameter $\Psi_B$ to undergo a discontinuous transition, and 
jump to the next plane-wave state, $\Psi_B \simeq \phi_1$. The fermionic order parameter then adjusts to this 
change, and is multiplied by the phase $e^{-4 i \theta}$. As a result,
\begin{eqnarray}
  \Psi_B = \phi_1, \,\, \Psi_P = \sqrt{1 - \frac {\tilde{\ell}} {x_P}} \phi_{-4} 
  + \sqrt{\frac {\tilde{\ell}} {x_P}} \phi_{-3}.
\end{eqnarray}
The same situation continues all the way up to $\ell = 1$. Then, the rest of the spectrum, for $\ell > 1$, is 
determined by Bloch's theorem \cite{FB}, in agreement with the numerical results of Fig.\,2.

Turning to Fig.\,\ref{fig:5}, the situation is more subtle. In a sense the role of the two components is reversed, 
since it is now energetically more favourable to keep the pairs -- and not the bosons -- in plane-wave states. 
There is one important difference, though, compared with the previous case. Although in Fig.\,\ref{fig:3} the 
slope of the dispersion relation at $\ell = x_P/2 = 0.05$ is continuous, in the present case, at $\ell \approx 
x_P/2 = 0.05$, it has a discontinuity, as seen in the upper plot of Fig.\,\ref{fig:5}. (This discontinuity 
appears also approximately at all the odd-integer multiples of $\ell = x_P = 0.05$, i.e., at $\ell \approx 
0.05, 0.15, 0.25$, etc.)

In order to understand this qualitatively, let us consider again the limit of weak interactions. For the trial 
states  
\begin{eqnarray}
\Psi_B = \sqrt{1 - \frac {\ell} {x_B}} \phi_0 + \sqrt{\frac {\ell} {x_B}} \phi_1, \,\, \Psi_P = \phi_0
\label{tr1}
\end{eqnarray}
the allowed values of $\ell$ are $0 \le \ell \le x_B$. Considering also the trial order parameters
\begin{eqnarray}
\Psi_B = \sqrt{\frac {x_P - \ell} {x_B}} \phi_{-1} + \sqrt{1 - \frac {x_P - \ell} {x_B}} \phi_0, 
\,\, \Psi_P = \phi_1
\label{tr2}
\end{eqnarray}
the allowed values of $\ell$ are $x_P-x_B \le \ell \le x_P$. Thus the common range of $\ell$ of the 
states of Eqs.\,(\ref{tr1}) and (\ref{tr2}) is $0 \le \ell \le x_P$. 

Evaluating the energy in the states of Eq.\,(\ref{tr1}) we find that, 
\begin{eqnarray}
 \frac E {N K} = \ell + 2 \epsilon_B \left[ \frac {\ell} {x_B} \left( 1 - \frac {\ell} {x_B} \right) \right].
\label{E1}
\end{eqnarray}
Similarly, for the states of Eq.\,(\ref{tr2}),
\begin{eqnarray}
 \frac E {N K} = 2 x_P - \ell 
 + 2 \epsilon_B \left[ \frac {x_P - \ell} {x_B} \left( 1 - \frac {x_P - \ell} {x_B} \right) \right].
\nonumber \\
\label{E2}
\end{eqnarray}
Figure 4 shows the two parabolas of Eqs.\,(\ref{E1}) and (\ref{E2}). There is a clear level crossing, which
leads to a discontinuous transition and also to the discontinuity in the slope of the dispersion relation. 
In the limit where the radius of the ring increases, with $n_B^0$ kept fixed, this takes place exactly at $\ell 
= x_P/2$. At this value of $\ell$ also the order parameter of the pairs undergoes a discontinuous 
transition from $\Psi_P \simeq \phi_0$ to the state $\Psi_P \simeq \phi_1$, up to $\ell = 3 x_P/2$, etc. Having 
understood the behaviour of the system at the interval $0 \le \ell < x_P$, the rest of the spectrum follows by 
center of mass excitation, according to Bloch's theorem, as in the case examined earlier.

\section{Effect of the mass imbalance between the bosonic atoms and the paired fermions}

Up to now we have assumed that $m_B = m_P$. We examine now the more general problem, where $m_B \neq m_P$ 
\cite{Anoshkin}. Let us consider the many-body wavefunction of the bosonic atoms and of the fermion pairs
in some interval of the total angular momentum $0 \le L_0 \le L_{\rm per}$,
\begin{eqnarray}
  \Psi_{L_0} = \Psi_{L_0}(\theta_1, \dots \theta_{N_B}, \varphi_1, \dots, \varphi_{N_P}).
\end{eqnarray}
Here the coordinates $\theta_i$, with $1 \le i \le N_B$, refer to the bosonic component, while $\varphi_i$, 
with $1 \le i \le N_P$, refer to the fermion pairs.

Motivated by the case of equal masses, let us investigate now the conditions which allow us to excite the center 
of mass of this two-component system. First of all, the center of mass coordinate $\Theta_{\rm CM}$ is
\begin{eqnarray}
  \Theta_{\rm CM} = \frac 1 {m_B N_B + m_P N_P} 
  \left(m_B \sum_{i=1}^{N_B} \theta_i + m_P \sum_{i=1}^{N_P} \varphi_i \right).
\nonumber \\
\end{eqnarray}
In order to achieve center of mass excitation with some integer multiple of $L_{\rm per}$, say $n L_{\rm per}$, 
we have to act with $e^{i n L_{\rm per} \Theta_{\rm CM}}$ on $\Psi_{L_0}$. This operation will give 
$\Psi_{L_0 + n L_{\rm per}}$.

In order to do that, and since we have to satisfy the periodic boundary conditions (without loss of generality 
we set $n=1$ for the moment), the two combinations which appear in the exponent, $L_{\rm per} m_B/(m_B N_B + 
m_P N_P)$ and $L_{\rm per} m_P/(m_B N_B + m_P N_P)$, have to be integers, say $p$ and $q$, respectively, i.e.,
\begin{eqnarray}
  \frac {L_{\rm per} m_B} {m_B N_B + m_P N_P} = p, \,\,\, \frac {L_{\rm per} m_P} {m_B N_B + m_P N_P} = q.
\end{eqnarray}
From the above two equations follows that
\begin{eqnarray}
 \frac {m_B} {m_P} = \frac p q
 \label{pcn}
\end{eqnarray}
and also
\begin{eqnarray}
  L_{\rm per} = p N_B + q N_P.
  \label{laper}
\end{eqnarray}
Therefore, in order to be able to excite the center of mass motion of the system, the ratio between 
the masses has to be a rational number. In addition, the period in $L$, $L_{\rm per}$, is no longer 
$N$, as in the symmetric model, i.e., $m_B = m_P$, but rather $L_{\rm per} = p N_A + q N_B$. Actually, 
the (smallest) period is the one that results from the values of $p$ and $q$, divided by their greatest 
common divisor. Apparently, for $p = q = 1$, we get the symmetric case, where $L_{\rm per} = N$.

We stress that the above results coincide with the ones in Sec.\,IV A, when Eq.\,(\ref{pcn}) is valid.
The difference is that in the case of phase separation, there is no restriction on the masses, while 
here the ratio between the masses has to be a rational number. The reason for this is the following. 
When we have phase separation, the density of the two components is sufficiently small at a certain 
spatial extent around the ring, which allows the phase of $\Psi_B$ and $\Psi_P$ to vary, satisfying 
the boundary conditions, without any effect on any physical observable. On the contrary, in the case 
of phase coexistence, this freedom in the phase match is no longer possible and the periodic boundary 
conditions require that Eq.\,(\ref{pcn}) holds.

Let us now turn to the energy spectrum. From the previous discussion the bosonic component takes an 
angular momentum $L_B = n p N_B$, while the pairs $L_P = n q N_P$ (we now take the more general 
case, with $n$ being any positive integer). Within the mean-field approximation, if the order parameters 
of the two components for $0 \le L_0 \le L_{\rm per}$ are expanded in the plane-wave states $\phi_m$
\begin{eqnarray}
 \Psi_B^0 = \sum_m c_m \phi_m, \,\,\, \Psi_P^0 = \sum_m d_m \phi_m,
\label{boo1}
\end{eqnarray}
then at any other interval with $n L_{\rm per} \le L \le (n+1) L_{\rm per}$,
\begin{eqnarray}
 \Psi_B^n = \sum_m c_{m} \phi_{m + n p}, \,\,\, \Psi_P^n = \sum_m d_m \phi_{m + n q}.
 \label{boo2}
\end{eqnarray}
It turns out that the total angular momentum $L_n$ in these states is, indeed, $L_n = L_0 + n L_{\rm per}$. 
Also, if $K_0$ is the total kinetic energy of the states of Eq.\,(\ref{boo1}), and $K_n$ is the total 
kinetic energy of the states of Eq.\,(\ref{boo2}), then
\begin{eqnarray}
 K_n - K_0 = \frac {\hbar^2} {2 M R^2} (L_{\rm per} n^2 + 2 L_0 n),
\end{eqnarray}
where $M = m_B/p = m_P/q$. The form of $K_n$ is 
\begin{eqnarray}
 K_n = \frac {\hbar^2} {2 M R^2} \left( n + \frac {L_0} {L_{\rm per}} \right)^2 L_{\rm per} = 
 \frac {\hbar^2} {2 M R^2} \frac {L_n^2} {L_{\rm per}},
\end{eqnarray}
and finally, the energy spectrum for the total energy $E$ per particle is
\begin{eqnarray}
  \frac E N = \frac {\hbar^2} {2 M R^2} \frac {L^2} {N L_{\rm per}} + e(L),
\label{blm}
\end{eqnarray}
where we have dropped the index $n$ in $L$. Here, $e(L)$ is a periodic function, with period 
$L_{\rm per}$. Finally, introducing ${\tilde K} = \hbar^2/(2 M R^2)$, Eq.\,(\ref{blm}) may be 
written as
\begin{eqnarray}
\frac {E} {N {\tilde K}} = \frac {\ell^2} {p x_B + q x_P} + \frac {e(\ell)} {\tilde K}.
\label{blmmm}
\end{eqnarray}
The first term on the right coincides with Eq.\,(\ref{kr}). Furthermore, for equal masses the above 
expression reduces to Eq.\,(\ref{bl}).

In Fig.\,5 we have considered an example of unequal masses, with $m_B/m_P = 1/3$, or $p = 1$ $q = 3$, 
and we have evaluated the dispersion relation, which is in agreement with Eq.\,(\ref{blmmm}). 
More specifically, we have considered the same $G_B$ and $G_{FB}$ as in Fig.\,2, $x_P=0.1$ and $x_B 
= 0.9$, and $\kappa=1/4$, i.e., in the BCS regime. As in the upper plots of Figs.\,2 and 3, we subtract 
again the energy due to the center of mass excitation, i.e., we plot $e(\ell)/{\tilde K}$, while in
the lower plot we also show how the angular momentum is distributed between the two species. From these 
plots we see the expected, exact, periodicity $L_{\rm per}/N = p x_B + q x_P = 1.2$ of $e(\ell)$. On top 
of that, we still have the quasi-periodicity, equal to $0.1$, set by the minority component, seen also 
in Figs.\,2, and 3, which was analysed in the previous section.

Clearly, in a real system, the ratio between the two masses is not a rational number in general. Still, 
even if this ratio is close to some rational number, one expects that the deviations from the derived 
spectrum to be perturbatively small \cite{Anoshkin}.

\begin{figure}
\includegraphics[scale=0.6]{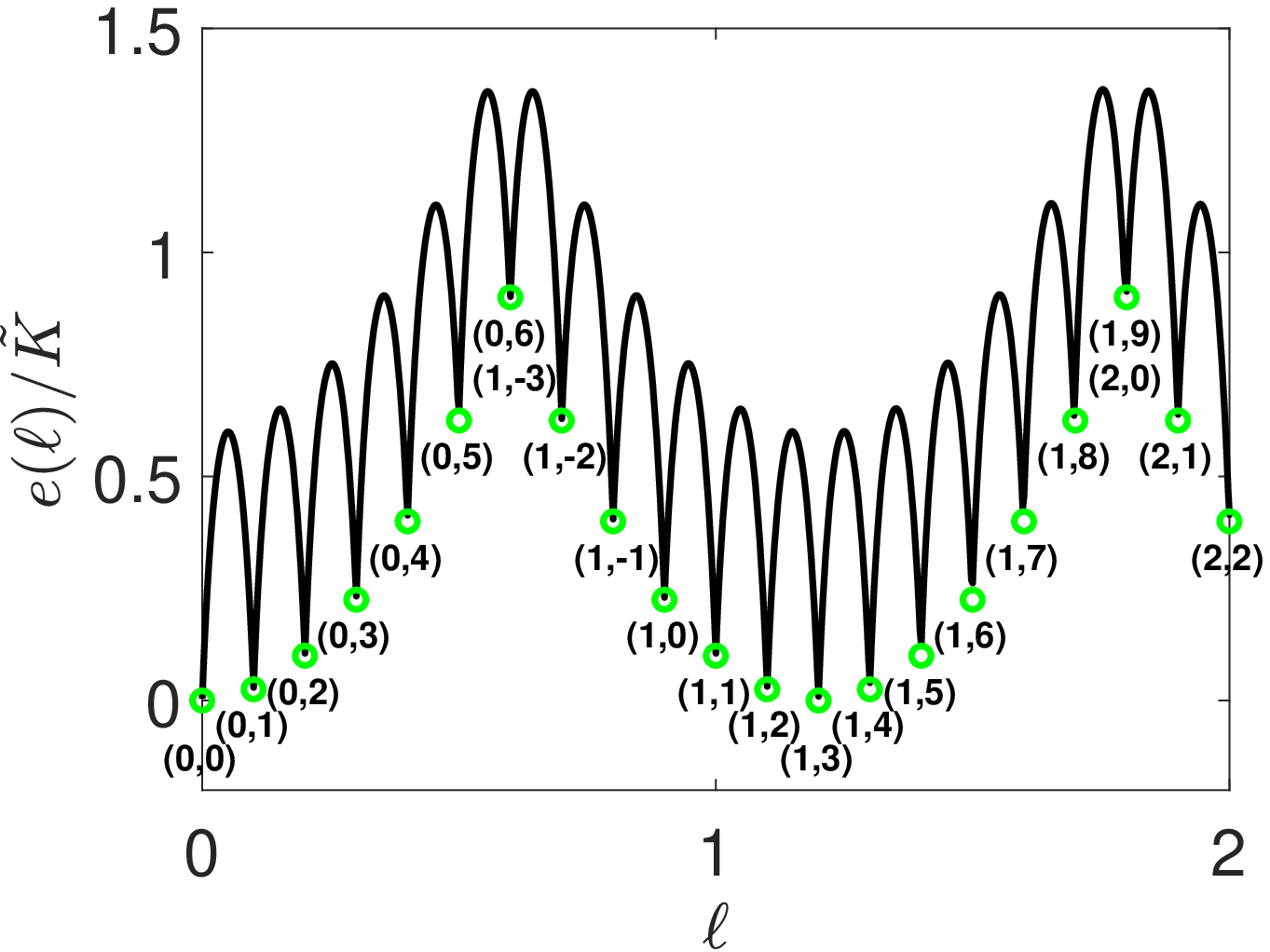}
\includegraphics[scale=0.6]{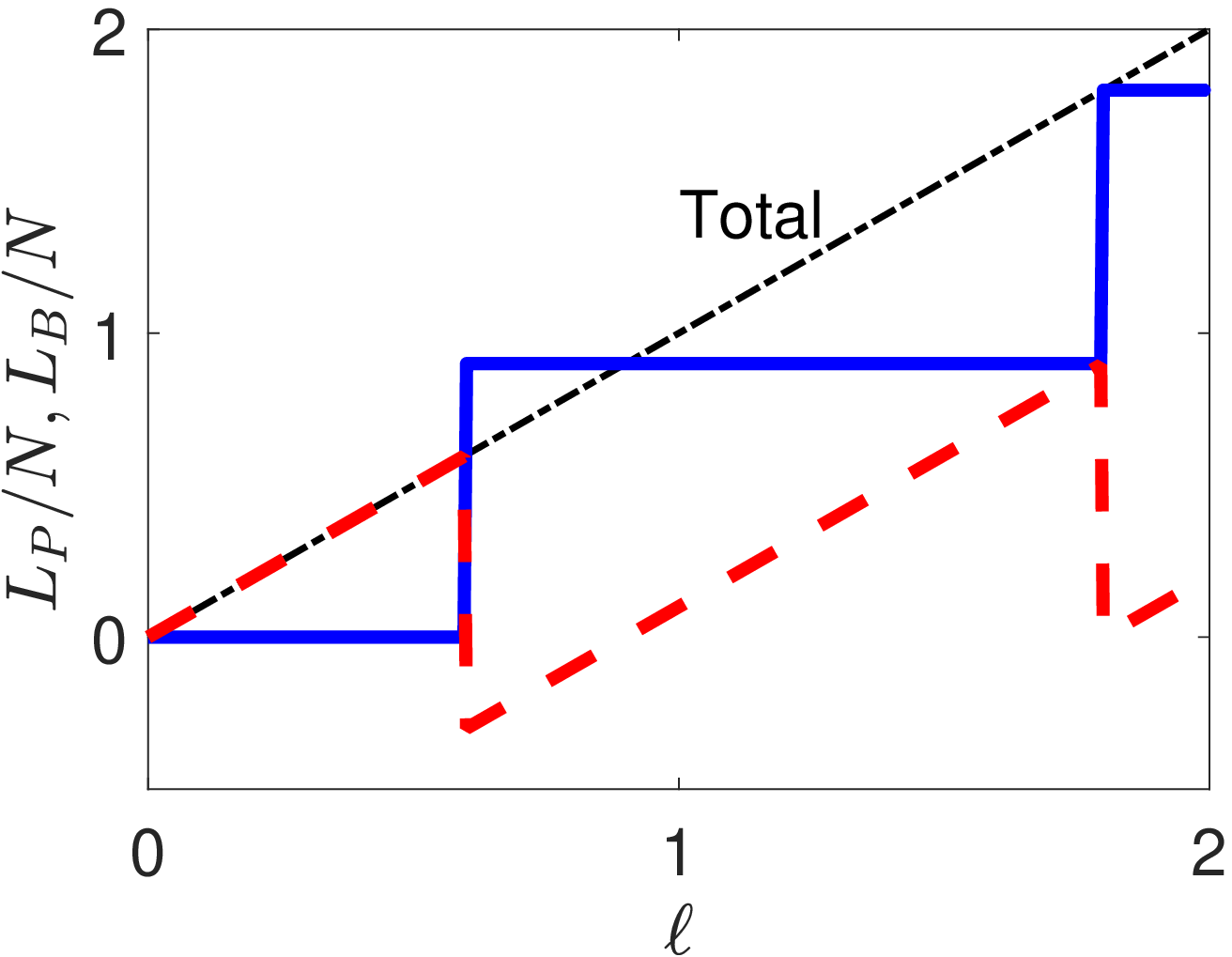} 
\caption{(Color online) The periodic function $e(\ell)$ (upper plot), and the distribution of the angular 
momentum between the two components (lower plot), in the regime of phase coexistence, for unequal boson 
and fermion-pair masses, $m_B = M$, $m_p = 3 M$. Here, the blue color (solid line) corresponds to the 
bosons, and the red color (dashed line) corresponds to the pairs. Also, $x_P=0.1$ and $x_B = 0.9$, in 
the BCS regime ($\kappa=1/4$), with $G_B$ and $G_{FB}$ being the same as in Fig.\,2. The periodicity 
$L_{\rm per}/N$, according to Eq.\,(\ref{laper}), is 1.2, as seen in the figures. The unit of energy 
${\tilde K}$ used here on the $y$ axis is $\hbar^2/(2 M R^2)$. The indices, e.g., $(0,0)$ denote the 
order parameter of the two superfluids at each local minimum, in the notation of Eq.\,(\ref{eq_PW}).}
\label{fig:difm} 
\end{figure}

\section{Summary and Conclusions}

In the present study we have considered a mixture of a Bose-Einstein condensate, with a paired fermionic superfluid,
at zero temperature. We have assumed that these two components are confined in a ring potential, as in a very
tight toroidal trapping potential. The periodic boundary conditions, combined with the degrees of freedom of the 
two superfluids, give rise to interesting effects. In the non-rotating, ground state of the system, clearly there 
are two phases that one may identify. In the one, the two components are distributed homogeneously around the 
ring, while in the other, the components separate spatially. 

The rotational response of the system, which is the main question that we have investigated here, depends crucially 
on the ground state. When the two components separate, the two components carry their angular momentum via center 
of mass excitation. 

The more interesting case is the one where the two components coexist uniformly (in the non-rotating, ground state). 
For small values of the angular momentum, we solved the problem via linearisation of the two coupled equations. 
This approach also allowed us to identify the nature of the sound-wave excitation of the system. For the more general 
problem, we solved the problem numerically. Interestingly enough, our results show that for a rather wide range of 
the parameters, and also for a relatively large population imbalance, the vast majority of the angular momentum is 
carried by the one component. This is not a surprise, since, for the rather strong non-linear terms we have considered 
(as in real experiments) it is energetically favourable for the components to maintain their homogeneous density 
distribution. As the angular momentum increases, the static component starts rotating, too. In certain cases (i.e., 
Figs.\,3 and 4), this is accomplished via discontinuous transitions. 

Another interesting consequence of the derived dispersion relation is related with the local minima, which show up 
at the integer multiples of the minority component. These minima may give rise to persistent currents. The high 
degree of tunability of these minima, which depend on the population imbalance, the strength of the nonlinear terms 
and the mass of the two components, is not only an interesting theoretical result, but it may also have technological 
applications.

In all of our displayed results we have assumed that the majority of the particles are the bosonic atoms. Still, 
the derived results are representative -- at least qualitatively -- of the phases that show up, also in the 
opposite limit, where the pairs is the dominant component. Actually, we argue that only in the special case where 
the populations of the two components are rather close to each other may the picture presented here be altered 
significantly (at least when the ratio between $L_{\rm per}$ and the population of the minority component is an 
integer multiple, as in the results considered in this study). In addition, according to Sec.\,V, no dramatic 
change occurs in the dispersion relation in the case of a mass imbalance, apart from the period of $e(\ell)$, 
provided that the mass ratio is a rational number, or close to it. 

Therefore, the present results are representative not only in terms of the population imbalance, but also in 
terms of the mass imbalance between the bosonic atoms and the fermionic pairs. We thus come to the conclusion 
that, despite the large parameter space that one has to cover in order to get the full picture, the present 
results cover a substantial fraction of the full phase diagram.

Compared with the problem of a bosonic mixture, the present problem has qualitative similarities. The main
difference lies in the nonlinear term that appears for the pairs of fermions. While in the Bose-Bose mixtures 
the energy per unit length scales quadratically with the density (as a result of the assumed s-wave collisions), 
here the nonlinear term that corresponds to the fermionic component has a stronger density dependence, which 
goes as the third power of the density. This dependence comes from the Fermi pressure of the fermionic atoms, 
which constitute the pairs, and in that sense it is of a very different nature. Interestingly enough, this term 
also resembles a three-body collision term in the Hamiltonian. 

It would definitely be interesting to investigate this problem also experimentally, in order to confirm 
the richness of the phases seen here. To make contact with experiment, for a radius $R = 100$ $\mu$m, $N = 
10^3$ atoms, for scattering lengths $a_B$ and $a_{FB}$ 100 \AA, for a transverse width of the torus
$1 \, \mu$m, and a population imbalance $N_P/N_B = 10$, one gets that $\epsilon_B \approx 10^3$, 
$\epsilon_{FB} \approx 10^2$, while $\epsilon_F \approx 10^2$. Furthermore, all the three energy scales 
$E_B/N$, $E_{FB}/N$, and $E_P/N$ are at least an order of magnitude smaller than the oscillator quantum 
of energy associated with the transverse degrees of freedom, and thus the motion of the atoms should be, 
to rather good degree, quasi-one-dimensional. Finally, the typical value of the speed of sound for these 
parameters is a few tens of mm/sec.

\section{Acknowledgments}

M.~\"{O}. acknowledges partial support from ORU-RR-2019. The authors wish to thank M. Magiropoulos and
J. Smyrnakis for useful discussions.

\appendix

\section{Derivation of the demixing condition} \label{app_sec_for_mixing_condition}

Let us assume that the order parameters have the sinusoidal form
\begin{equation}
\Psi_{B}  = \frac 1 {\sqrt{2 \pi R}} (c_{0} + 2 c_{1} \cos \theta),  \, \, 
\Psi_{P}  = \frac 1 {\sqrt{2 \pi R}} (d_{0} + 2 d_{1} \cos \theta),  
\label{ModelForMixing}
\end{equation}
where $c_0^2+2c_1^2 = 1$ and $d_0^2+2d_1^2 = 1$. The total energy of the system in the states of 
Eq.~(\ref{ModelForMixing}) is the following quadratic form,
\begin{widetext}
\begin{equation}
E = 2 N_B \frac {\hbar^2} {2 m_B R^2} c_1^2 + 2 N_P \frac {\hbar^2} {2 m_P R^2} d_1^2  
+   2 N_B^2 \frac{G_B}{\pi R} c_1^2 - 4 N_B N_P \frac{G_{FB}}{\pi R} c_1 d_1
+ 2 N_P^3 \frac{G_P}{\pi^2 R^2} d_1^2.  
\end{equation}
\end{widetext}
Minimizing the resulting quadratic equation, and demanding that the determinant of the linear system vanishes 
we get that
\begin{eqnarray}
\left( \frac{\hbar^2}{2 m_B R^2}  +   \frac{G_B N_B  }{\pi R}  \right) 
\left( \frac{\hbar^2}{2 m_P R^2}  +   \frac{G_P N_P^2}{\pi^2 R^2}  \right) =
\nonumber \\
=  \frac{G_{FB}^2 N_B  N_P}{\pi^2 R^2}, 
\end{eqnarray}
which gives the boundary for the phase coexistence/separation of Eq.~(\ref{CriticalMixingDemixing}). 
We stress that in the limit where the kinetic-energy terms above are negligible, the above condition is 
equivalent to Eqs.~(18) and~(19) in~\cite{AdhikariPRA2007}.

\section{Derivation of the Bogoliubov spectrum} \label{BSp}

We assume small deviations of the order parameters from the homogeneous solution $\Psi_B^0 = \phi_0$ and 
$\Psi_P^0 = \phi_0$, i.e., $\Psi_B(\theta, t) = \Psi_B^0 + \delta \Psi_B (\theta, t)$ and $\Psi_P(\theta, t) 
= \Psi_P^0 + \delta \Psi_P (\theta, t)$. Linearising the following two coupled time-dependent equations, 
\begin{eqnarray}
 i \hbar \frac {\partial \Psi_B} {\partial t} &=& 
 \left[ -\frac{\hbar^2 \partial_{\theta \theta} }{2 m_B R^2}  
 +  G_B N_B |\Psi_B|^2 +  G_{FB} N_P |\Psi_P|^2  \right] \Psi_B, 
 \nonumber \\
 i \hbar \frac {\partial \Psi_P} {\partial t} &=& 
 \left[ -\frac{\hbar^2 \partial_{\theta \theta}} {2 m_P R^2}   
 + G_P N_P^2 |\Psi_P|^4 +  G_{FB} N_B |\Psi_B|^2 \right] \Psi_P ,  
\nonumber \\
 \label{sys1t}
\end{eqnarray}
we get that
\begin{widetext}
\begin{eqnarray}
 i \hbar \frac {\partial (\delta \Psi_B - \delta \Psi_B^*)} {\partial t} =
 - \frac {\hbar^2} {2 m_B R^2} \frac {\partial^2 (\delta \Psi_B + \delta \Psi_B^*)} {\partial \theta^2} +
 2 G_B n_B^0 (\delta \Psi_B + \delta \Psi_B^*) + 2 G_{FB} \sqrt{n_B^0 n_P^0} (\delta \Psi_B + \delta \Psi_B^*),
\end{eqnarray}
\end{widetext}
and also
\begin{eqnarray}
 i \hbar \frac {\partial (\delta \Psi_B + \delta \Psi_B^*)} {\partial t} =
 - \frac {\hbar^2} {2 m_B R^2} \frac {\partial^2 (\delta \Psi_B - \delta \Psi_B^*)} {\partial \theta^2}.
\end{eqnarray}
Similar equations also hold for $\delta \Psi_P$. Assuming plane-wave solutions, and demanding that the 
resulting homogeneous system of two equations with two unknowns has a non-trivial solution, we get the
usual condition, which then leads to Eq.\,(\ref{ds}), 
\begin{widetext}
\begin{eqnarray}
\begin{vmatrix}
\begin{pmatrix}
\hbar^2 k^4/(2 m_B R^2) + 2 G_B n_B^0 k^2 - 2 m_B \omega^2 R^2  
&  2 G_{FB} \sqrt{n_B^0 n_P^0} k^2  \\ 
2 G_{FB} \sqrt{n_B^0 n_P^0} k^2 & \hbar^2 k^4/(2 m_P R^2) + 4 G_P (n_P^0)^2 m^2 - 2 m_P \omega^2 R^2   
\end{pmatrix}
\end{vmatrix}
= 0.
\label{matrix}
\end{eqnarray}
\end{widetext}

\end{document}